\documentclass[referee]{raa}            

\usepackage{graphicx,time}             
\usepackage{epstopdf}
\usepackage{natbib}
\usepackage{amssymb}
\usepackage{multirow,multicol}
\usepackage{accents}

\bibpunct{(}{)}{;}{a}{}{,}
\usepackage[pagebackref=true]{hyperref}
\hypersetup{pdftitle = The title of my PDF, pdfauthor = My name, pdfsubject= The subject, pdfkeywords = keyword1 keyword2 keyword3}
\hypersetup{colorlinks = true, linkcolor = green, anchorcolor = red, citecolor = blue, filecolor = red, pagecolor = red, urlcolor = red}

\newcommand*{\pbar}[1]{\accentset{(-)}{#1}}

\begin{document}

\title{A global analysis for searching neutrinos associated with the black holes merging gravitational wave events
  \,$^*$
  \footnotetext{$*$ Supported by the National Natural Science Foundation of China.}
}

\volnopage{Vol.0 (20xx) No.0, 000--000}      
\setcounter{page}{1}          

\author{Yu-Zi Yang\inst{1}, Jia-Jie Ling\inst{1}$^{\dagger}$, Wei Wang\inst{1}
  \and Zhao-Kan Cheng \inst{2}
  \footnotetext{$\dagger$ Corresponding author}
}


\institute{Sun Yat-sen University,
  No. 135, Xingang Xi Road, Guangzhou, Guangdong 510275, China;
  {\it yangyz6@mail2.sysu.edu.cn, lingjj5@mail.sysu.edu.cn, wangw223@mail.sysu.edu.cn}\\
  \and
  Harbin Institute of Technology,
  No. 92, Xidazhi Street, Harbin, Heilongjiang 150001, China;
  {\it 14B911018@hit.edu.cn}\\
}

\date{Received~~2009 month day; accepted~~2009~~month day}

\abstract{Several neutrino observatories have searched
for coincident neutrino signals associated with gravitational waves induced by the merging of two black holes.
No statistically significant neutrino signal in excess of background level was observed.
These experiments use different neutrino detection technologies and are sensitive to various neutrino types.
A combined analysis was performed on the KamLAND, Super-Kamiokande and Borexino experimental data
with a frequentist statistical approach to achieve a global picture of the associated neutrino fluence.
Both monochromatic and Fermi-Dirac neutrino spectra were assumed in the calculation.
The final results are consistent with null neutrino signals associated with a binary of black holes merging process.
The derived 90\% confidence level upper limits on the fluence and luminosity of various neutrino types are presented for neutrino energy less than 110 MeV.
}

\authorrunning{Y.-Z. Yang et al.}            
\titlerunning{A global analysis on black holes merger GW events}  

\maketitle

%
%
\section{Introduction}           
\label{sect:intro}

After the detections of the gravitational waves (GW) released from black hole-black hole (BH-BH) mergers in 2015 by the Advanced Laster Interferometer Gravitational-wave Observation (LIGO) (\citealt{adLIGO:2015al}), many efforts have been made to search for the coincident signals in astronomical observations. No significant counterpart except a coincident gamma-ray burst which was observed by the Fermi Telescope (\citealt{Connaughton:2016umz}) at ~0.4 seconds after GW150914 observation (\citealt{Abbott:2016a}). In general, there is no theory of neutrino generation associated with BH-BH mergers. However, some physical phenomena such as gravitational radiation, the synthesis of heavy elements, and short gamma ray bursts are closely tied to the emission of neutrinos from BH accretion disk systems (\citealt{Caballero:2015cpa}).
In the accretion disk system theory, neutrinos play a crucial role of setting the initial electron fraction of the outflow matter (\citealt{r-Process:2008}). Neutrino annihilation can increase the prevalence of gamma-ray bursts (\citealt{grbneu:1999}). The appearance of a gamma ray burst implies the accretion disk system theory may also work for BH-BH mergers. The search for GWs, gamma ray bursts and neutrinos can open a new window for multi-messenger research of BH-BH mergers. This will lead to a more complete understanding of cosmic processes through a combination of information from different probes, and increase search sensitivity over any single detection method (\citealt{Adrian-Martinez:2016xgn}).

Two observed gravitational wave events in 2015 were both generated by the coalescence of binary BHs into a single BH. The first event, GW150914 (\citealt{Abbott:2016b}), was observed on September 14, 2015. Two initial BHs of $36_{-4}^{+5} M_{\odot}$ and $29_{-4}^{+4} M_{\odot}$, at $410_{-180}^{+160}$ Mpc luminosity distance from the earth, merged to a final BH of $62_{-4}^{+4} M_{\odot}$ with $3_{-0.5}^{+0.5} M_{\odot}$ radiated in gravitational waves. The second one, GW151226 (\citealt{Abbott:2016c}), was observed on December 26, 2015. Masses of the two initial BHs were $14.2_{-3.7}^{+8.3} M_{\odot}$ and $7.5_{-2.3}^{+2.3} M_{\odot}$, and the mass of the final BH was $20.8_{-1.7}^{+6.1} M_{\odot}$. The source was located at a distance of $440_{-190}^{+180}$ Mpc from the earth.


\section{Experimental Searches}
\label{sect:Neu}

Several neutrino observatories, including Super-Kamiokande, KamLAND, Borexino, ANTARES, IceCube and Pierre Auger, have tried to search for the neutrino signals associated with the two black hole merger events.
Since neutrino generation mechanism from the merger of two BHs is still unclear, all of those experiments conservatively choose a $\pm$500 seconds coincidence time window centered around the gravitational wave observation time. More details about those experimental searches are introduced in the following paragraphs.

The KamLAND experiment (\citealt{Gando:2016zhq}) is located under 2700 meter water equivalent (m.w.e.) of vertical rock, below Mt.Ikenoyam in Gifu prefecture of Japan. The detector consists of a stainless-steel sphere and EVOH/nylon outer balloon. The outer balloon encloses 1 kton of liquid scintillator (LS). During the GW150914, a mini-balloon was placed at the center of the detector to search for neutrinoless double beta decay. KamLAND focuses on the search for electron antineutrinos which are mainly detected through inverse beta-decay (IBD) reaction: $\bar{\nu}_{e} + p \rightarrow e^{+} + n$. The coincidence of the prompt ($e^+$ --ionization and annihilation) and delayed ($n$ -- capture on Hydrogen) signals efficiently suppresses the backgrounds. The prompt signal has a strong bound with the $\overline{\nu}_e$ energy, where $E_{\rm Prompt} \approx E_{\nu} - 0.8$ MeV. However, there are two major drawbacks. One is the loss of low energy $\bar{\nu}_{e}$ information below IBD interaction threshold around 1.8 MeV. Another is the obscuration of the incoming neutrino's direction information due to the recoil neutrons diffusion. During the GW150914 and GW151226 periods, no neutrino candidates were detected by KamLAND.

Borexino (\citealt{Agostini:2017pfa}) is also a LS experiment located underground at 3400 m.w.e. in Gran Sasso Laboratory, Italy. Besides the IBD detection channel, Borexino is also capable to detect neutrinos via electron scattering (ES) thanks for its high purity scintillator with extremely low radioactive backgrounds. In comparison with the IBD method, ES does not have the energy threshold and is sensitive to all flavors of active neutrinos ($\pbar{\nu}_{e}$, $\pbar{\nu}_{\mu}$ and $\pbar{\nu}_{\tau}$). However, it is still challenging for a LS detector to determine the incoming neutrino direction by reconstructing the scattered electron with scintillator light. In the two GW periods, no neutrino signal was found in the IBD channel and one signal candidate was detected in the ES channel.

The Super-Kamiokande (Super-K) experiment (\citealt{Abe:2016jwn}) has a large 50 kton water Cherenkov detector located at 2700 m.w.e. underground in Kamioka Japan. After a neutrino interacts inside of the detector, the Cherenkov ring pattern reconstruction can identify the final product of the charged particles from which they can infer the neutrino's direction, flavor and energy. Neutrinos with energies from 3.5 MeV to 79.5 MeV are categorized as the ``low energy data sample'' which typically solar neutrinos and supernova relic neutrino. The data set with energy above 100 MeV is typically used to study atmospheric neutrinos and the proton decay search. Because of its relatively ``high energy'' detection threshold, the observable signals in the Super-K detector originate only from charged particles, which are positrons and electrons for IBD and ES interactions respectively. For that reason, Super-K is unable to separate these two kinds of interactions. Four neutrino candidates were observed by Super-K for GW150914 and no candidate was observed for GW151226.

The IceCube and ANTARES experiments (\citealt{Adrian-Martinez:2016xgn}) have also searched for coincident neutrino candidates in their recorded data. They are primarily sensitive to neutrinos with $\gg$ GeV energies. During the GW150914 and GW151226 periods, three neutrino candidates were found for each GW event, however these signals do not satisfy the spatial and temporal requirements.

\section{Global analysis}
\label{sect:ana}

As introduced above, with different detection technologies, different neutrino experiments are sensitive to different neutrino flavors and energy ranges. Performing a global analysis on those experimental data can produce a full picture of the black hole merger coincident neutrino searches. Based on limited available information from the publications, a global analysis was done with the ``low energy'' data samples from the Super-Kamiokande, KamLAND and Borexino experiments, which are sensitive to neutrino energy less than 110 MeV.

Before performing a combined analysis, it is essential to understand each experiment's data and reproduce their results. Given the reported low statistical signal candidates and background estimation, a maximum likelihood ratio method based on Poisson statistics is adopted to estimate the corresponding neutrino signal.
\begin{equation}
  \chi^{2}(\mu) = 2\left[ (\mu + n_{\rm bg}) - n_{\rm obs} + n_{\rm obs} \cdot ln\frac{n_{\rm obs}}{\mu + n_{\rm bg}}\right],
\end{equation}
\noindent where $\mu$ is the number of expected neutrino signals from the BH-BH merger in the coincidence window, $n_{\rm bg}$ is the estimated background, and $n_{\rm obs}$ is the number of observed coincidence candidates.

Since all of those experiments did not observe any statistically significant neutrino signals, the estimated neutrino signals are consistent with zero. More importantly, as reported by those experiments, we also calculate $N_{90}$, which is upper limit of the neutrino signal at 90\% confidence level. Because of low event statistics, the traditional confidence interval setting method based on Wilk's theorem (\citealt{Wilks:1938dza}) is inaccurate. The Feldman-Cousins method (\citealt{Feldman:1997qc}) based on toy Monte Carlo simulation is chosen to estimate the upper limit of the detected neutrino signal. Specifically, for each value of $\mu$, 90\% of the toy Monte Carlo events have smaller $\chi^{2}(\mu) - \chi^{2}(\mu_{\rm best})$ values than $\Delta \chi^{2}_{90}(\mu)$, where $\mu_{\rm best}$ is the best-fit value. The value of $\Delta \chi^{2}_{\rm data}(\mu)$ was also calculated for each $\mu$ using the actual number of observed candidates in the experimental data. $N_{90}$ equals the maximal values of $\mu$ when $\Delta \chi^{2}_{\rm data}(\mu) \leq \Delta \chi^{2}_{90}(\mu)$.

As shown in Table~\ref{Tab:neutrino-information}, our reproduced neutrino signal upper limits are consistent with the reported experimental results.

\begin{table}
  \begin{center}
    \caption[]{Experimental results of KamLAND, Borexino and Super-K neutrino observatories on GW150914 (GW151226) period, including the detection method, neutrino energy ranges, signal candidates, background estimations, and for both the published and reproduced 90\% C.L. upper limits on signal $N_{90}$.}\label{Tab:neutrino-information}
    \begin{tabular}{ccccccc}
      \hline\hline\noalign{\smallskip}
      Experiment & Channel & Energy (MeV) & $\nu$ Candidates & Backgrounds & $N_{90}$ (published) & $N_{90}$ (Reproduced)                   \\
      \hline\noalign{\smallskip}
      KamLAND & IBD & 1.8-110 & 0 (0) &  0.18 (0.02) & 2.26 (2.41) & 2.25 (2.41)  \\
      \hline
      \multirow{2}{*}{Borexino} & IBD & 1.8-75 & 0 (0) & $\sim$0 ($\sim$0) & 2.44 (2.44) & 2.43 (2.43)\\
      & ES & 0.4-15 & 0 (1) & 1.68 (1.72) & N/A &1.3 (2.8)\\
      \hline
      Super-K & IBD/ES & 3.5-79.5  & 4 (0) & 2.90 (2.90) & 5.41 (2.30) & 5.92 (1.0) \\
      \noalign{\smallskip}\hline\hline
    \end{tabular}
  \end{center}
  \tablecomments{\textwidth}{Super-K used Bayesian statistical methods to calculate the upper limits of neutrino signals (\citealt{Abe:2016jwn}). We recalculate their results with the frequentist Feldmen-Cousins approach. Borexino doesn't report $N_{90}$ result on electron scattering channel.}
\end{table}

The upper limit on the signal $N_{90}$ can be converted into the upper limit of neutrino fluence, $F_{\rm UL}$, which is the neutrino flux at the earth. The calculation formula is given by
\begin{equation}
  F_{\rm UL} = \frac{N_{90}}{N_{T}\int{\phi(E_{\nu})\sigma(E_{\nu})\epsilon(E_{\nu})dE_{\nu}}},
\label{eq:fluence}
\end{equation}
\noindent where $N_{T}$ stands for the total number of target nuclei, which are protons for the IBD interactions and electrons for ES. $\phi(E_{\nu})$ is the normalized neutrino energy spectrum, $\epsilon(E_{\nu})$ is the total detection efficiency, including the time window selection efficiency, detection efficiency, etc. $\sigma(E_{\nu})$ is the neutrino interaction cross-section. The IBD cross section is taken from~\citealt{Strumia:2003zx}; while the ES differential cross section is shown in Equation~\ref{Elastic} (\citealt{Giunti:2007ry}).

\begin{equation}\label{Elastic}
\frac{d\sigma(E_{\nu},T_{e})}{d T_{e}}=\frac{2G^{2}_{F}m_{e}^{2}}{\pi}[g_{1}^{2}+g_{2}^{2}(1-\frac{T_{e}}{E_{\nu}})^{2}-g_{1}g_{2}\frac{m_{e}T_{e}}{E_{\nu}^{2}}]
\end{equation}
\noindent where $G_{F}$ is the Fermi constant, $m_{e}$ is the electron mass, and $T_{e}$ is the kinetic energy of the recoil electron. $E_{\nu}$ stands for neutrino energy.  $g_{1}, g_{2}$ are the coupling constants: $g_{1}^{(\nu_{e})}=g_{2}^{(\bar{\nu}_{e})}=\frac{1}{2}+\sin^{2}\theta_{W}$ and $g_{2}^{(\nu_{e})}=g_{1}^{(\bar{\nu}_{e})}=\sin^{2}\theta_{W}$ for electron type neutrinos (anti-neutrinos), where $\theta_{W}$ is the Weinberg angle;
$g_{1}^{(\nu_{\mu,\tau})}=g_{2}^{(\bar{\nu}_{\mu,\tau})}=-\frac{1}{2}+\sin^{2}\theta_{W}$ and $g_{2}^{(\nu_{\mu,\tau})} = g_{1}^{(\bar{\nu}_{\mu,\tau})} = \sin^{2}\theta_{W}$ for muon or tau type neutrinos (antineutrinos).

For a global analysis, the expected number of neutrino signals at various experiments are different due to differing target mass, interaction cross section and detection efficiencies. Instead, we can directly introduce neutrino fluence, $F$, in the total $\chi^{2}(F)$ as shown in Equation~\ref{totalchi2}.
\begin{equation}
  \chi^{2}(F) = \sum_{i}^{\rm exps}{\chi^{2}_{i}(F)} = 2 \sum_{i}^{\rm exps} {\left[(F \cdot N_{T} \cdot \sigma_{{\rm eff},i} + n_{{\rm bg},i})-n_{{\rm obs},i}+ n_{{\rm obs},i} \cdot ln \frac{n_{{\rm obs},i}}{F \cdot N_{T} \cdot \sigma_{{\rm eff},i} + n_{{\rm bg}, i}}\right]},
  \label{totalchi2}
  \end{equation}
\noindent where $\sigma_{{\rm eff}, i}$, the effective cross section embeds the detector efficiency of the $i$-th experiment, is defined as:
\begin{equation}
\sigma_{{\rm eff}, i} = \int{\phi(E_{\nu})\sigma(E_{\nu})\epsilon(E_{\nu})dE_{\nu}}.
\label{P1}
\end{equation}
\noindent Hence the upper limits for neutrino fluence $F_{UL}$ can also be computed using the Feldman-Cousins method, similar to the calculation of the upper limit for the neutrino signal $N_{90}$, as described previously. The systematic correlations among those measurements are negligible except for the theoretical model of neutrino production from BH-BH mergers, which are independent of experimental measurement.

Since the neutrino energy spectrum $\phi(E_{\nu})$ is unclear, the upper limits for neutrino fluence were calculated using two hypotheses: monochromatic and Fermi-Dirac distributions.

\subsection{Monochromatic energy spectrum}
The simplest hypothesis for the neutrino energy spectrum assumes all neutrinos have the same energy and follow a $\delta (E_{\nu})$ distribution. This assumption provides the most conservative (or the largest) upper limits on neutrino fluence at a given neutrino energy $E_{\nu}$.

\begin{figure}[ht]
  \centering
  \includegraphics[width=7.2cm, angle=0]{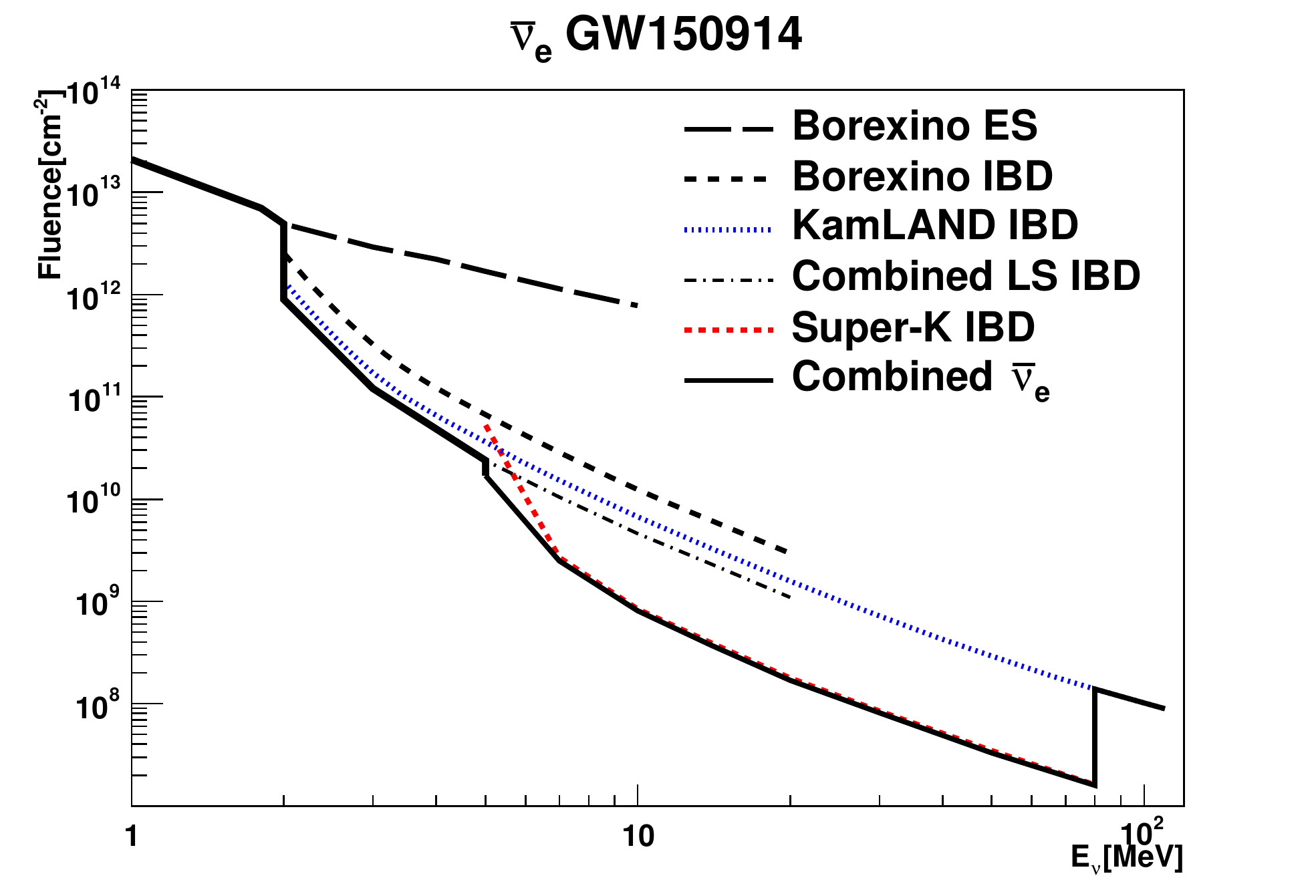}
  \includegraphics[width=7.2cm, angle=0]{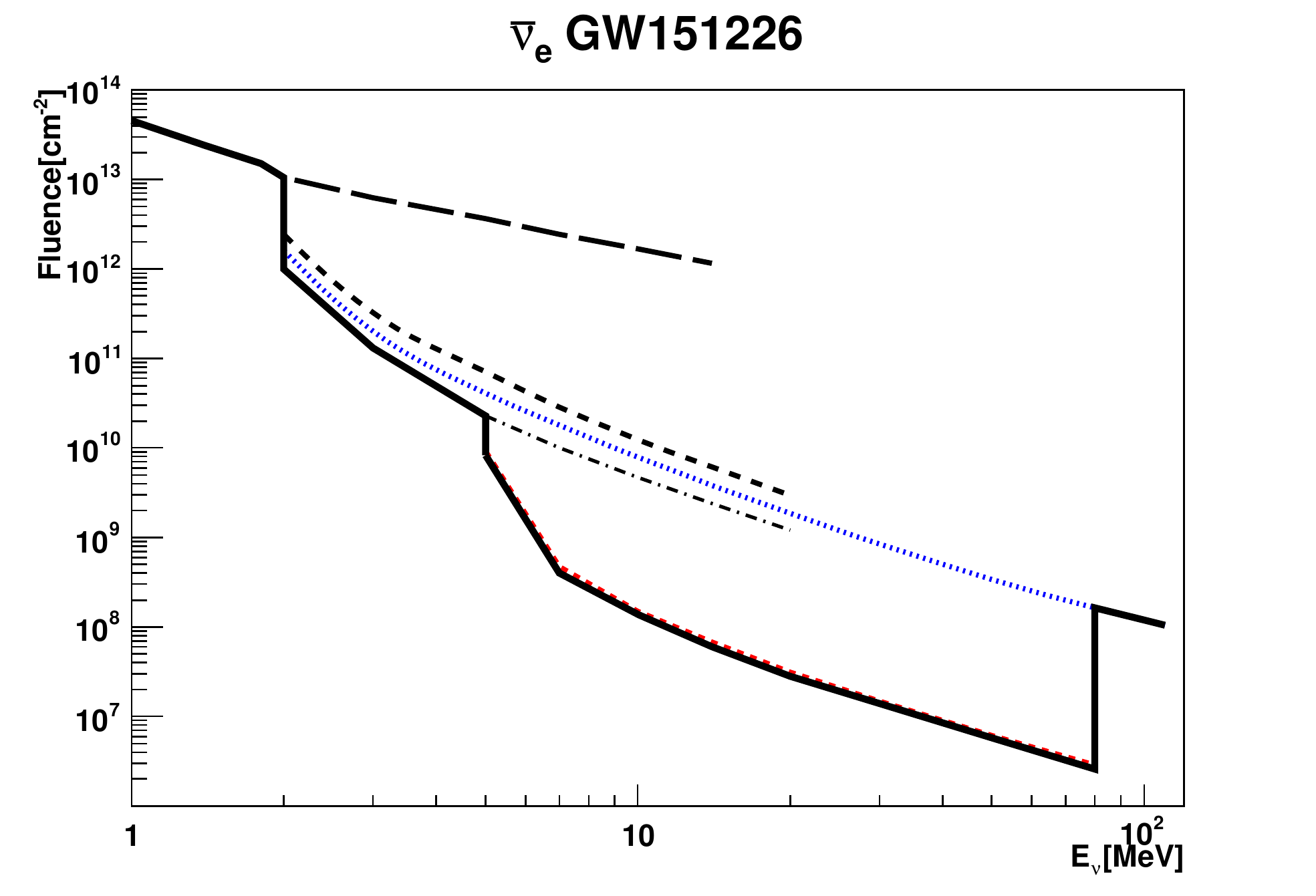}
  \caption{The 90\% C.L. upper limits on $\bar{\nu}_{e}$ fluence from GW150914 and GW151226 BH-BH mergers with respect to the monochromatic neutrino energy $E_{\nu}$. The dotted line represents the combined analysis result for KamLAND and Borexino while solid dashed line represents the global analysis result based on all three experiments.}
  \label{antinue}
\end{figure}

Two global analysis were done. One only included the two largest LS experiments (KamLAND and Borexino), because LS detectors have high sensitivities to low energy $\bar{\nu}_{e}$ detection. The other included  all three experiments. The 90\% C.L. upper limits on $\bar{\nu}_{e}$ neutrino fluence with respect to neutrino energy for GW150914 and GW151226 BH-BH mergers are shown in Figure~\ref{antinue}. The combined upper limits of $\bar{\nu}_e$ fluence are from $1.0 \times 10^{12}$ cm$^{-2}$ to $1.6 \times 10^{7}$ cm$^{-2}$ (from $1.0 \times 10^{12}$ cm$^{-2}$ to $2.9 \times 10^{6}$ cm$^{-2}$) for GW150914 (GW151226) at the energy range ($1.8$ MeV $<E_{\nu}<79.5$ MeV). Since the IBD cross section is proportional to E$^{2}_{\nu}$, hence the corresponding neutrino fluence is inversely proportional to E$^{2}_{\nu}$ according to equation~\ref{eq:fluence}. For extremely low energy neutrinos ($E_{\nu}<1.8$ MeV), the resulting sensitivity only comes from the Borexino experiment through ES interaction channel; For low energy neutrinos ($1.8$ MeV $<E_{\nu}<5$ MeV), LS experiments, especially KamLAND, contribute mainly to the sensitivity due to larger IBD cross section; while for neutrino energy ($5.0$ MeV $<E_{\nu}<79.5$ MeV), Super-K dominates the sensitivity due to its huge target mass.

\begin{figure}[ht]
  \centering
   \includegraphics[width=7.2cm, angle=0]{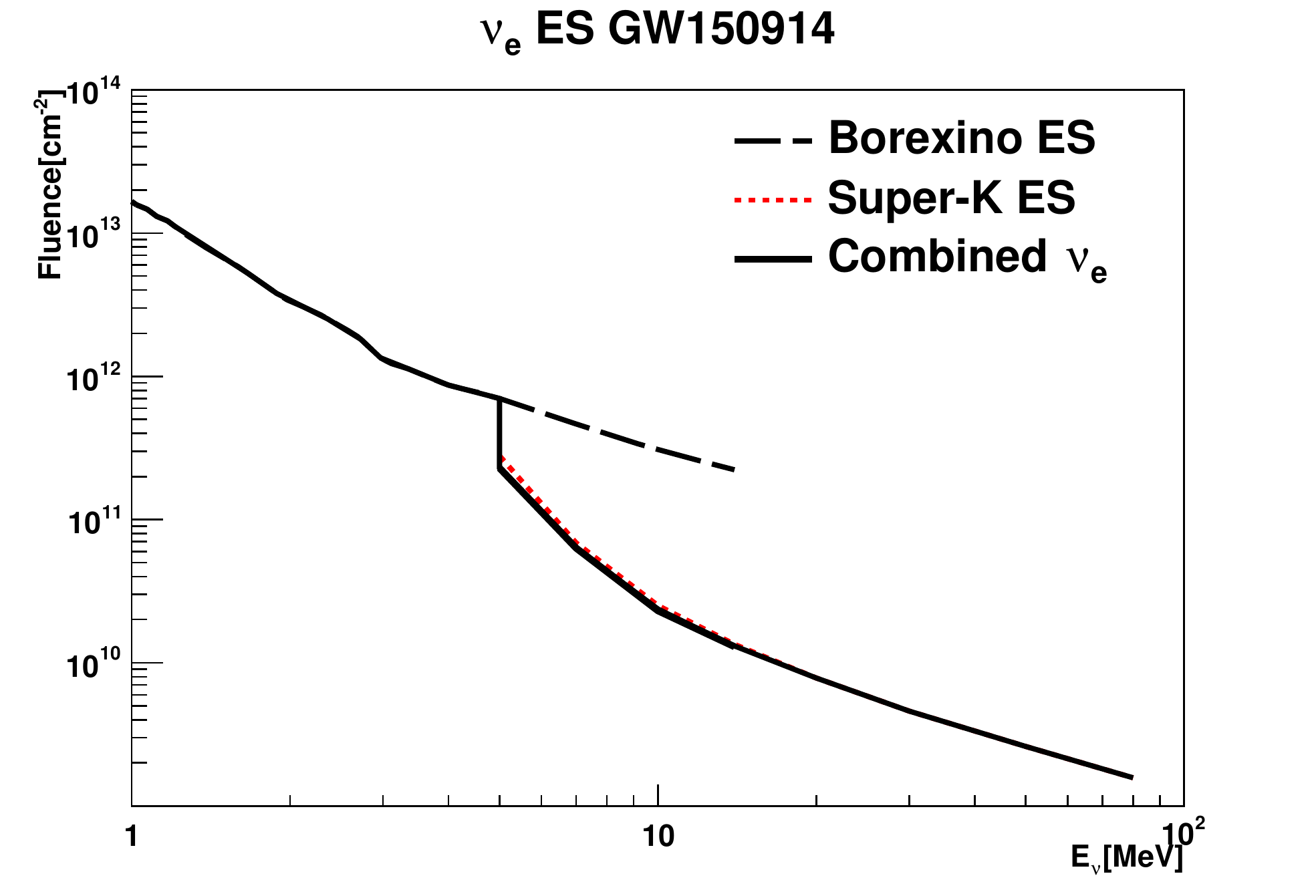}
   \includegraphics[width=7.2cm, angle=0]{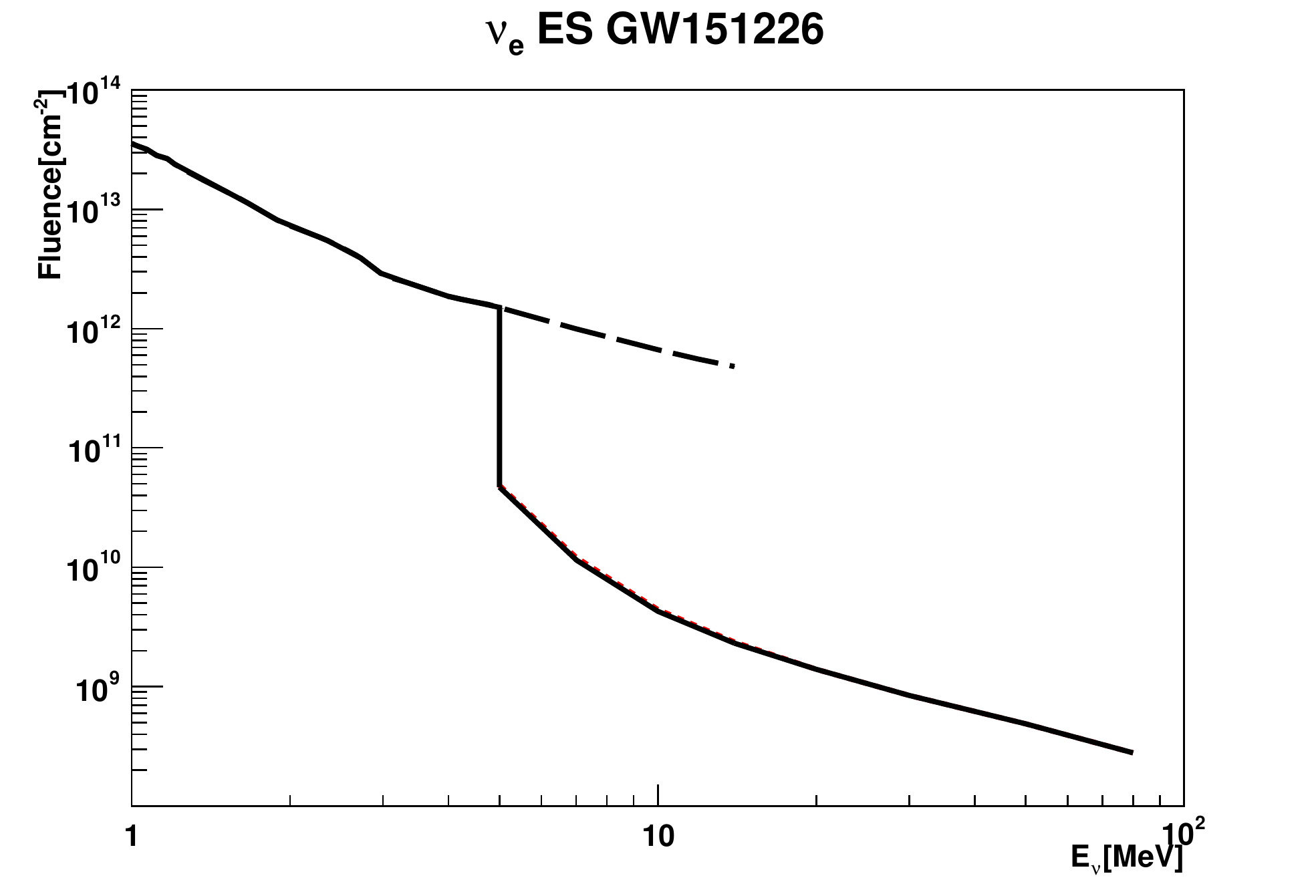} \\
   \includegraphics[width=7.2cm, angle=0]{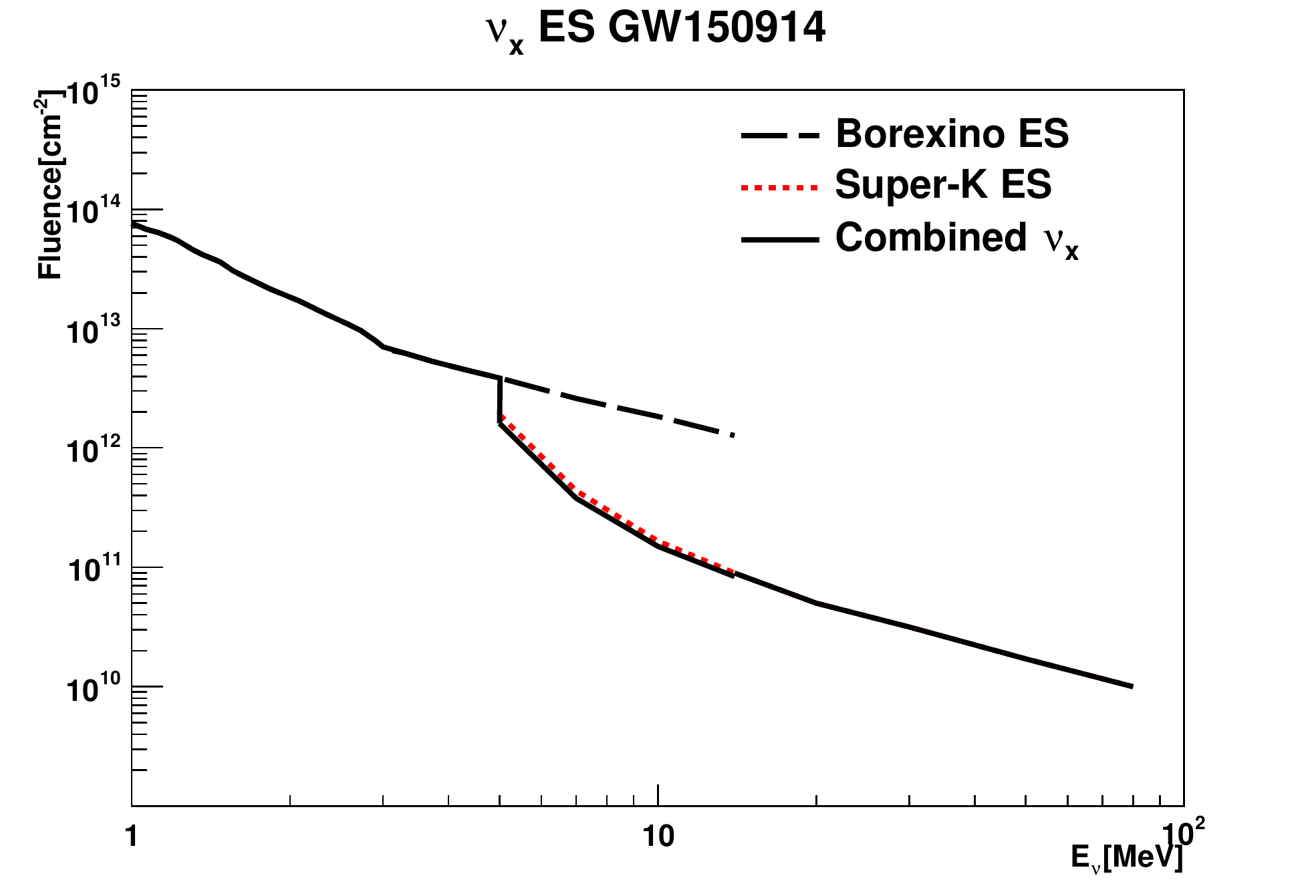}
   \includegraphics[width=7.2cm, angle=0]{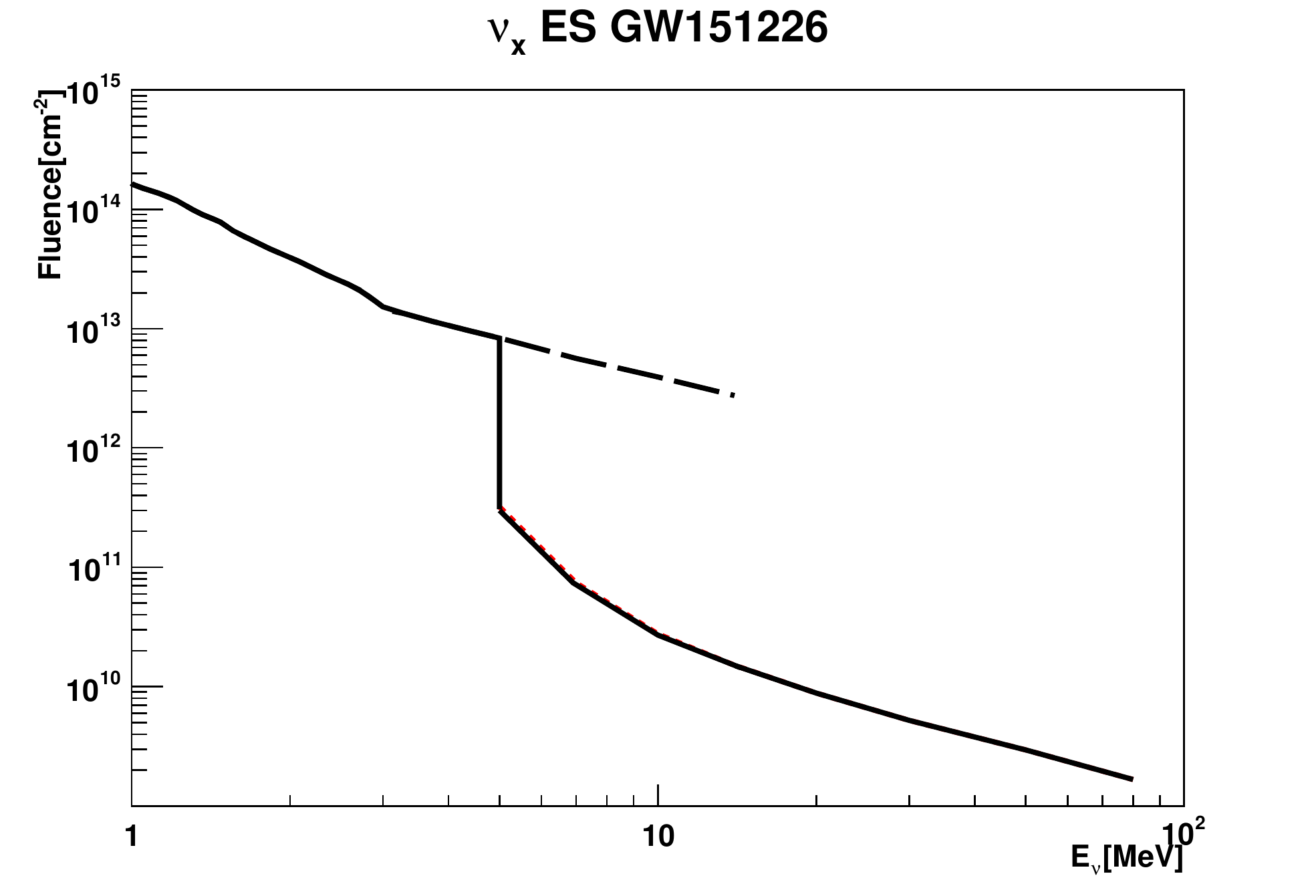}
   \caption{The 90\% C.L. upper limits on $\nu_{e}$ and $\nu_{x}$ fluence from GW150914 and GW151226 BH-BH mergers as a function of the monochromatic neutrino energy $E_{\nu}$.}
   \label{nuenux}
\end{figure}

The fluence of other types of neutrinos, $\nu_{e}$ and $\nu_{x}$ ($\nu_{\mu}$ and $\nu_{\tau}$), can only be deduced from the ES channel. As shown in Figure~\ref{nuenux}, the combined 90\% C.L. upper limits for $\nu_e$, $\nu_x$ fluence are from $1.7\times10^{13}$cm$^{-2}$ to $1.6 \times 10^9$cm$^{-2}$ and from $7.7 \times 10^{13}$ to $1.0 \times 10^{10}$cm$^{-2}$ for GW150914 at the energy range of $1.0$ MeV $<E_{\nu}<79.5$ MeV. For GW151226, the obtained upper limits are from $3.6 \times 10^{13}$cm$^{-2}$ to $2.8 \times 10^8$cm$^{-2}$ and from $1.6 \times 10^{14}$cm$^{-2}$ to $1.7 \times 10^9$cm$^{-2}$ for $\nu_{e}$ and $\nu_{x}$. According to equation~\ref{Elastic}, the ratio of $\sigma_{\nu_e}$/$\sigma_{\bar{\nu}_e}$ $\simeq 2.5$, so the upper limits of $\nu_e$ and $\bar{\nu}_e$ fluence from Borexino are similar. While $\sigma_{\nu_e,ES}$ is about two orders of magnitude lower than $\sigma_{\bar{\nu}_e,IBD}$, therefore the resulting upper limit of $\nu_e$ would be two orders of magnitude higher than $\bar{\nu}_e$'s in the Super-K experiment.

\subsection{Fermi-Dirac energy spectrum}
During a binary black hole merging process, the neutrino released may obey the Fermi-Dirac energy distribution with zero chemical potential $\eta$=0:
\begin{equation}\label{Fermi-Dirac}
  \phi_{\rm Fermi-Dirac}(E_{\nu}) = Norm \cdot \frac{E^{2}_{\nu}}{1+e^{E_{\nu}/T-\eta}},
\end{equation}
\noindent where $Norm$ stands for a normalization factor, $T$ is the effective neutrino temperature, which is set to be 5 MeV as the nominal value.

\begin{table}[ht]
\begin{center}
  \caption[]{90\% C.L. upper limits on $\nu_{e}$, $\bar{\nu}_{e}$ and ${\nu}_{x}$ fluence from Borexino, KamLAND, Super-K experiments and the combined results for GW150914 (GW151226).}\label{Tab:fluence}
  \begin{tabular}{cccc}
    \hline \hline\noalign{\smallskip}
    \multirow{2}{*}{Experiment} & \multicolumn{3}{c}{90\% C.L. F$_{\rm UL}$ (cm$^{-2}$)} \\
    \cline{2-4}
     & $\nu_{e}$ & $\bar{\nu}_{e}$ & ${\nu}_{x}$ \\
    \hline
    KamLAND & - & $2.0 \times 10^{9}$ ($2.4 \times 10^{9}$) & - \\
    \hline
    Borexino & $3.4 \times 10^{11}$ ($7.4 \times 10^{11}$) & $2.7 \times 10^{9}$ ($2.7 \times 10^{9}$) &  $1.9 \times 10^{12}$ ($4.1 \times 10^{12}$) \\
    \hline
    Super-K & $1.2 \times 10^{10}$ ($2.0 \times 10^{9}$) & $2.6 \times 10^{8}$ ($4.0 \times 10^{7}$) & $7.2 \times 10^{10}$ ($1.2 \times 10^{10}$) \\
    \hline
    Combined & $1.2 \times 10^{10}$ ($2.0 \times 10^{9}$) & $2.2 \times 10^{8}$ ($3.7 \times 10^{7}$) & $7.1 \times 10^{10}$ ($1.2 \times 10^{10}$) \\
    \noalign{\smallskip}\hline\hline
  \end{tabular}
\end{center}
\end{table}

90\% C.L. upper limits on the integrated $\nu_{e}$, $\bar{\nu}_{e}$ and ${\nu}_{x}$ fluence and luminosity obtained from KamLAND, Borexino and Super-K experiments and their combined results are shown in Table~\ref{Tab:fluence} and Table~\ref{Tab:luminosity}, where the luminosity $L$ is the total released energy taken by neutrinos during a BH-BH merging process. It can be calculated using $L = F \cdot 4 \cdot \pi \cdot D_{\rm gw}^{2} \cdot <\bar{E}>$ equation, where $D_{\rm gw}$ is the distance from the GW source to the earth, $<\bar{E}>$ is the average neutrino energy which equals to 3.15$\cdot$T. Since the uncertainties of the measured distances for those GW events are quite large, luminosity is written as a function of the true distance to the source

\begin{equation}
  L_{\rm GW150914} = L_{0}\left(\frac{D_{\rm gw}}{410 {\rm Mpc}} \right)^{2} {\rm ergs}
\end{equation}
\noindent and
\begin{equation}
  L_{\rm GW151226} = L_{0}\left(\frac{D_{\rm gw}}{440 {\rm Mpc}} \right)^{2} {\rm ergs}
\end{equation}
\begin{table}[h]
\begin{center}
  \caption[]{90\% C.L. upper limits on $\nu_{e}$, $\bar{\nu}_{e}$ and ${\nu}_{x}$ neutrino luminosity from Borexino, KamLAND, Super-K experiments and the combined results for GW150914 (GW151226) without oscillation. }\label{Tab:luminosity}
  \begin{tabular}{cccc}
    \hline \hline\noalign{\smallskip}
    \multirow{2}{*}{Experiment} & \multicolumn{3}{c}{90\% C.L. L$_{0 \rm UL}$ (ergs)} \\
    \cline{2-4}
    & $\nu_{e}$ & $\bar{\nu}_{e}$ & ${\nu}_{x}$ \\
    \hline
    KamLAND & - & $1.0 \times 10^{60}$ ($1.4 \times 10^{60}$) & - \\
    \hline
    Borexino & $1.7 \times 10^{62}$ ($4.3 \times 10^{62}$) & $1.4 \times 10^{60}$ ($1.6 \times 10^{60}$) & $9.6 \times 10^{62}$ ($2.4 \times 10^{63}$) \\
    \hline
    Super-K & $6.1 \times 10^{60}$ ($1.2 \times 10^{60}$) & $1.3 \times 10^{59}$ ($2.4 \times 10^{58}$) & $3.6 \times 10^{61}$ ($6.9 \times 10^{60}$) \\
    \hline
    Combined & $6.1 \times 10^{60}$ ($1.2 \times 10^{60}$) & $1.2 \times 10^{59}$ ( $2.2\times 10^{58}$) & $3.6 \times 10^{61}$ ($6.9 \times 10^{60}$) \\
    \noalign{\smallskip}\hline\hline
  \end{tabular}
\end{center}
\end{table}

\begin{figure}[ht]
   \centering
   \includegraphics[width=7.2cm, angle=0]{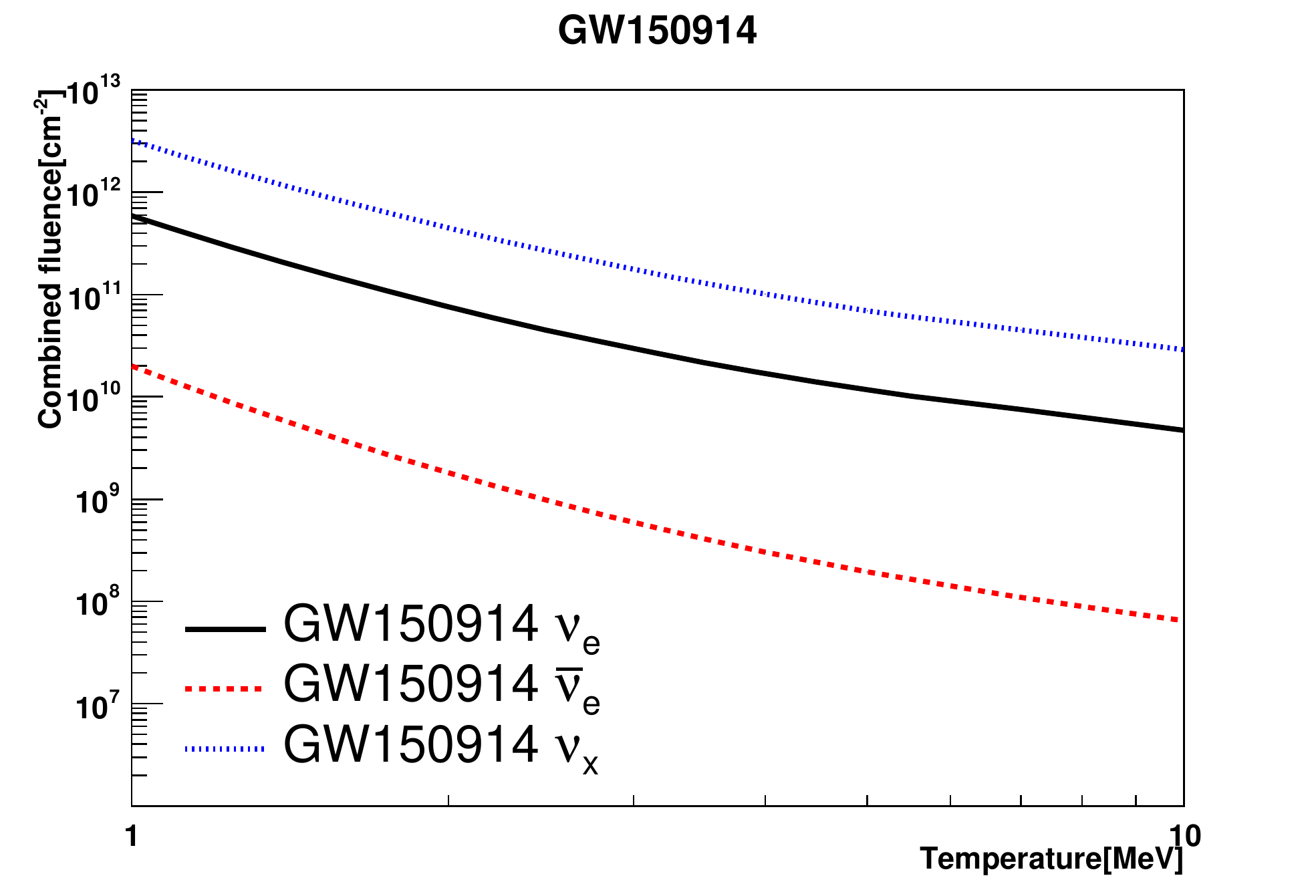}
   \includegraphics[width=7.2cm, angle=0]{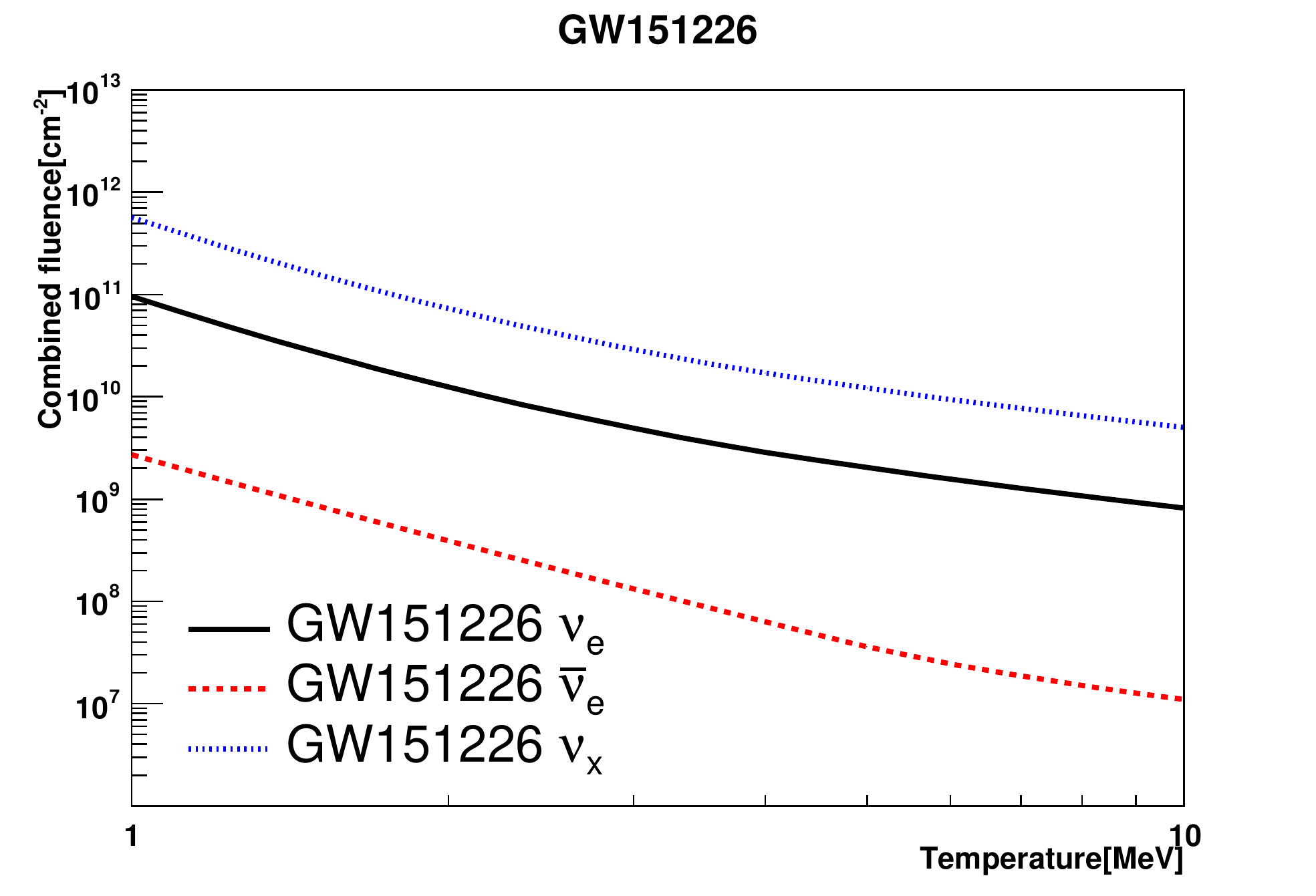}
   \caption{90\% C.L. upper limits on neutrino fluence for GW150914 and GW151226 for various types based on the Fermi-Dirac energy spectrum with respect to effective neutrino temperature $T$.}
   \label{fig:temp}
   \end{figure}

As shown in Figure~\ref{fig:temp}, 90\% C.L. upper limits on the combined $\nu_{e}$, $\bar{\nu}_{e}$ and ${\nu}_{x}$ neutrino fluence obtained from the Fermi-Dirac energy distribution are shown with respect to the effective neutrino temperature $T$. As $T$ increases, the averaged neutrino energy will also increase, which results a larger neutrino interaction cross section. As a result, the derived neutrino fluence will decrease with respect to $T$ as shown in Equation~\ref{eq:fluence}.

\section{Discussion}
\label{sect:discussion}
In this analysis each GW event is calculated separately. If the mechanism of neutrino generation from a BH-BH merger process is the same, these GW events can be analyzed together with some assumptions on their origin. The neutrino fluence may proportional to the mass of two merger BHs or the mass of a remnant relic, and its velocity of rotation. It is not clear whether the release of neutrinos occurs at the merger phase or cooling phase. That will also affect the arrival time of the neutrino signals relative to the corresponding gravitational wave signal. In addition, at this stage the uncertainties associated with the black hole masses and distances from the earth are quite large. More precise measurements with more statistical analysis in the future will be very helpful for the further study in this topic. It will help us to understand the physics behind BH-BH mergers, as well as the underlying neutrino physics.

\section{Conclusions}
\label{sect:conclusion}
A global analysis was performed on experimental data from the KamLAND, Super-Kamiokande and Borexino experiments to search for neutrinos associated with GW150914 and GW151226 events with a frequentist statistical approach. The final results are consistent with null neutrino signals associated with a binary of black holes merging process. For GW150914, the obtained 90\% C.L. upper limits on $\bar{\nu}_e$, $\nu_e$ and $\nu_x$ fluence are from $2.1 \times 10^{13}$cm$^{-2}$ to $1.6 \times 10^7$cm$^{-2}$, from $1.7 \times 10^{13}$cm$^{-2}$ to $1.6 \times 10^9$cm$^{-2}$ and from $7.7 \times 10^{13}$cm$^{-2}$ to $1.0 \times 10^{10}$cm$^{-2}$ in the energy range of $1.0$ MeV $<E_{\nu}<79.5$ MeV assuming monochromatic energy spectrum; While assuming Fermi-Dirac energy spectrum with an effective temerature of 5 MeV, the combined 90\% C.L. upper limits on $\nu_e$, $\bar{\nu}_e$ and $\nu_x$ fluence are $1.2 \times 10^{10}$ cm$^{-2}$, $2.2 \times 10^8$ cm$^{-2}$ and $ 7.1 \times 10^{10}$ cm$^{-2}$. Similar results are also obtained for GW151226 with slightly better upper limits.


\begin{acknowledgements}
We would like to thank Neill Raper, Mikhail Smirnov of Sun Yat-sen University, Ruiting Xie of Imperial College London for polishing the text. We also would like to thank A.P. Rong-Feng Shen of Sun Yat-sen University for the discussion of BH-BH merger. This work is funded by the National Natural Science Foundation of China (NSFC) under No.11080922.
\end{acknowledgements}

\bibliographystyle{raa}
\bibliography{msRAA-2018-0065}

%
%
%

\end{document}